\documentclass[pra,amsmath,notitlepage,aps,reprint]{revtex4-2}

\usepackage{graphicx}
\usepackage{dcolumn}
\usepackage{bm}
\usepackage{hyperref}
\usepackage{setspace}
\usepackage{amsfonts}

\begin{document}

\title{
The \texorpdfstring{$O(n\to\infty)$}{O(n->infinity)} Rotor Model and the Quantum Spherical Model on Graphs
}

\author{Nikita Titov}
\email{nikita.titov@units.it}
\affiliation{Department of Physics, University of Trieste, Strada Costiera 11, 34151 Trieste, Italy}
 
\author{Andrea Trombettoni}
\email{atrombettoni@units.it}
\affiliation{Department of Physics, University of Trieste, Strada Costiera 11, 34151 Trieste, Italy}

\date{\today}

\begin{abstract}
We show that the large $n$ limit of the $O(n)$ quantum rotor model defined on a general graph has the same critical behavior as the corresponding quantum spherical model and that the critical exponents depend solely on the spectral dimension $d_s$ of the graph. To this end, we employ a classical to quantum mapping and use known results for the large $n$ limit of the classical $O(n)$ model on graphs. Away from the critical point, we discuss the interplay between the Laplacian and the Adjacency matrix in the whole parameter plane of the quantum Hamiltonian. These results allow us to paint the full picture of the $O(n)$ quantum rotor model on graphs in the large $n$ limit.
\end{abstract}
\maketitle

\section{Introduction}
A wide range of problems and systems across different fields are most naturally formulated on graphs, to describe connections or interactions between a set of variables, such as spins, particles, or data defined on their nodes. As a consequence, one has to address both the graph structure and the model defined on it, which are in principle independent.

In many physical contexts, one chooses to minimize the intricacy of the former by considering translational invariant systems, which can be diagonalized by Fourier transformation \cite{cardy1996scaling,zee2010quantum}. However, this is not possible in non-translational invariant contexts. In this case, graph theory allows one to characterize and describe the underlying topology of these networks in various ways 
\cite{harary1969graph,bondy1976graph,barthelemy2022spatial}.

Choosing different models on these graphs leads to distinct behavior with applications in a plethora of areas and physical systems, such as ultracold atoms  \cite{buonsante2002bose,buccheri2016holographic}, 
superconducting Josephson junctions \cite{sodano2006inhomogenous,silvestrini2007topology,lucci2022quantum}, but also random walks \cite{burioni2001random} and combinatorial optimization problems \cite{goemans1995improved,andrew2014ising}.

Critical phenomena on graphs have been extensively studied, as they offer a way to characterize universal aspects of different classes of networks \cite{burioni1996universal,millan2021complex}. In this context, a key result is that for the classical spherical model it is possible to exactly determinate the critical exponents, which depend only on the spectral dimension $d_s$ of the graph \cite{cassi1999spherical}. This is owed to the remarkable simplicity of the spherical model \cite{berlin1952spherical}, which is a classical model combining continuous spins, described by gaussian variables, and a constraint on their average length. On translational invariant lattices, the spherical model was shown to be equivalent to the large $n$ limit of the classical $O(n)$ model \cite{stanley1969exact}, which gave a geometric interpretation and physical relevance to it. This equivalence, however, is not present when considering graphs breaking translational invariance \cite{titov2025laplacian, dantchev2014casimir, khoruzhenko1989large}. In these cases, the single global constraint on the spin length has to be exchanged for many local constraints, making analytical solutions rare. Nevertheless, it was shown in \cite{burioni2000limit} that the singular part of the free energies of both the spherical model and $O(n)$ model in the large $n$ limit are equal even on more general graphs. 

Quantum versions of the spherical model are well known in the literature \cite{vojta1996quantum,gracia2004quantum} and are usually obtained by promoting the spin variables to operators and introducing canonical momenta. They proved to be a paradigmatic example of models exhibiting a quantum phase transition, with critical exponents that can be determined exactly, again due to the inherent simplicity of the spherical model \cite{vojta1996quantum}. As in the classical case, also the quantum spherical model on translationally invariant lattices is equivalent to another model in the large $n$ limit. For the Hamiltonian considered in \cite{vojta1996quantum}, which is also the one studied in this paper, this turns out to be the $O(n)$ quantum rotor model, {i.e.}, the large $n$ limit of the $O(n)$ quantum rotor model is provided by the quantum spherical model. Quantum rotor models appear as effective descriptions in many different physical contexts \cite{sachdev2011quantum}. We make note specifically of the relation between the Bose Hubbard model at large filling and the $O(2)$ rotor model \cite{polak2007quantum, anglin2001exact}. The spherical model as the large $n$ limit cannot capture effects related to the compactness of the phase variable, but can still be a viable starting point for systems defined on graphs.

In this paper, our aim is to clarify and discuss the relation between the $O(n)$ quantum rotor and the quantum spherical models on graphs. Specifically, a first natural question is whether their critical behavior is the same in the large $n$ limit and whether the spectral dimension of the underlying graph determines it. We address this question by using a classical to quantum mapping to generalize the results obtained in \cite{cassi1999spherical,burioni2000limit}. We also address a second question concerning the large $n$ limit of the O(n) quantum rotor model away from criticality and the nature of the resulting effective quantum model. We show that, while at the critical point the spherical model is governed by the Laplacian matrix connecting the modes or sites of the graph, this is no longer the case away from criticality.

We then study the behavior of the Lagrange multipliers of the $O(n)$ quantum rotor model in the large $n$ limit and show that their behavior exhibits an interplay between Laplacian matrix and Adjacency matrix. The obtained findings are complementary to the corresponding results for the large $n$ limit of the classical $O(n)$ models away from the critical point \cite{titov2025laplacian} and give a complete picture of the large $n$ limit of classical and quantum $O(n)$ models at and away from the criticality.

The remainder of this paper is organized as follows. In Sec. \ref{section2} we briefly give definitions of the crucial graph-theoretic matrices and summarize known results for the classical $O(n)$ and spherical models. In Sec. \ref{section3} we perform the large $n$ limit of the $O(n)$ quantum rotor model on graphs and afterwards demonstrate how the Laplacian and Adjacency matrix arise from it in Sec. \ref{section4}. Sec. \ref{section5} is devoted to the generalization of the classical results concerning the critical behavior.

\section{The Classical \texorpdfstring{$O(n\to\infty)$}{O(n->infinity)} and Spherical Models on Graphs}\label{section2}
A graph $\mathcal{G}$ is a set of points and connections between them. The properties of physical models on graphs are strongly dependent on the spectral behavior of two matrices ubiquitous in graph theory \cite{harary1969graph,bondy1976graph,merris1994laplacian},
Firstly, the Adjacency matrix $A$, defined as
\begin{equation}
    A_{ij}=\begin{cases}
        1&\text{if sites $i$ and $j$ are connected,}\\
        0&\text{otherwise,}
    \end{cases}
\end{equation}
and secondly the Laplacian matrix $L$:
\begin{equation}
    L_{ij}=\delta_{ij}z_i-A_{ij},
\end{equation}
where $z_i=\sum_j A_{ij}$ is the coordination number of site $i$ ({\it i.e.}, the number of its nearest-neighbours). The Laplacian is positive semidefinite with a single zero eigenvalue (for a connected graph) and the corresponding ground state is homogeneous on \emph{any} graph. 

The Adjacency matrix on the other hand, has a ground state which favors localization around vertices with large coordination number. This difference is not present on regular lattices since there $z_i=z$ for all $i$ and the spectra are equal up to a constant shift. On more general graphs there is no simple connection between them, which gives rise to distinct low-energy phenomena such as the localization of particles \cite{burioni2001bose,brunelli2004topology,burioni2005random,millan2021complex}. It is therefore of importance to use the "correct" matrix when defining the model on graphs with inhomogeneous coordination number, with "correct" referring to the appropriate modelization of the system at hand. For example, if bosons may jump between nearest-neighbour sites of a graph with no other constraints, then the Adjacency matrix has to be used in the microscopic (Bose-Hubbard, in this case) model.

Next, we recall some results for the classical 
$O(n)$ and spherical models, which will be required in the following and serve as a starting point for the generalization to the quantum case. 

The classical $O(n)$ model describes $n$-component vectors $\mathbf{S}_i$ with normalized length:
\begin{equation}
    \mathbf{S}_i^2=\sum_{\alpha=1}^nS_{i,\alpha}=n,
\end{equation}
where the site index is $i$ and the components are denoted by $\alpha$. The normalization is in principle arbitrary, but we choose it in such away to obtain a meaningful large $n$ limit. On a graph with Adjacency matrix $A$ the Hamiltonian is given by:
\begin{equation}
    -\beta H=K\sum_{i,j=1}^NA_{ij}\mathbf{S}_i\cdot \mathbf{S}_j,\label{On_def}
\end{equation}
where $K=\beta J>0$ is a dimensionless, ferromagnetic coupling ($\beta=1/T$). 

In translational invariant lattices, the classical result by Stanley \cite{stanley1968spherical} is that the large $n$ limit of the 
$O(n)$ model \eqref{On_def} is provided by the spherical model, having as Hamiltonian:
\begin{equation}
    -\beta H = K \sum_{i,j=1}^N S_i A_{ij} S_j,
\end{equation}
with the constraint $\langle\sum_{i=1}^{N} S_i^2\rangle=N$ and the $S_i$ are now scalar variables.

On a general graph, the large $n$ limit can be taken the same way as on a regular lattice \cite{stanley1968spherical} with the crucial difference that it is necessary to introduce as many Lagrange multipliers as sites. The Hamiltonian in this limit reads \cite{titov2025laplacian}:
\begin{equation}
    -\beta H = \sum_{i,j=1}^NS_i(KA_{ij}-\delta_{ij}\lambda_i)S_j \equiv -\sum_{i,j=1}^NS_iM_{ij}S_j,\label{hamiltonian_large_n},
\end{equation}
where again $S_i$ are scalar variables and the Lagrange multipliers are determined such that the constraints $\langle S_i^2\rangle=1,\; \text{for } i=1,\dots,N$ are fulfilled. The number of constraints diverges in the thermodynamic limit which makes this model generally not easy to deal with. Nevertheless, there have been results over the years concerning the critical behavior \cite{cassi1999spherical,burioni2000limit}, but also away from the critical point \cite{titov2025laplacian}. We stress that the model \eqref{hamiltonian_large_n} is the large $n$ limit of the classical $O(n)$ model both at and away from the critical point, provided that the $\lambda_i$ have been determined.

In \cite{burioni2000limit} it has been shown that the critical behavior of \eqref{hamiltonian_large_n} is the same as of a generalization of the spherical model to graphs. Here there is a single constraint $\langle\sum_iz_i\ S_i^2\rangle=N$, which is enforced by a single Lagrange multiplier $\lambda$. The Hamiltonian of this spherical model reads:
\begin{equation}
    -\beta H = \sum_{i,j=1}^NS_i(KA_{ij}-\lambda\delta_{ij}z_i)S_j, \label{hamiltonian_cassi}
\end{equation}
It is important to note only the singular parts of the free energies $\beta f=-\log{Z}/N$, where $Z$ is the partition function, of both models \eqref{hamiltonian_large_n} and \eqref{hamiltonian_cassi} are equal to each other close to a phase transition. This however does not mean that $\lambda_i=\lambda z_i$ at the critical point. The form of the constraint in \eqref{hamiltonian_cassi} makes it possible to determine the critical exponents exactly because it relates directly to the Laplacian matrix. In other words, as soon as the critical properties are considered, the large-distance properties of the $O(n)$ models are given by the behaviour of the Laplacian matrix, not the Adjacency one. In \cite{cassi1999spherical} it was shown that the critical exponents of \eqref{hamiltonian_cassi} are determined solely by the so-called \emph{spectral dimension} $d_s$ of the graph ({i.e.,} the spectral dimension of the Laplacian matrix) and follow from those of the standard spherical model with the identification $d\to d_s$.

The behavior of $\lambda_i$ in the actual large $n$ limit on a graph has been recently analyzed in \cite{titov2025laplacian}, where it was shown that it can be understood as an interplay between the Laplacian and Adjacency matrices. At low temperatures the Lagrange multipliers force $M$ to approach the Laplacian matrix such that its spectrum determines the full free energy. At high temperatures, on the other hand, the Lagrange multipliers become all equal such that one obtains the Adjacency matrix spectrum (shifted by a large constant). 

These three distinct results concerning the critical behavior, but also a general feature of the large $n$ limit are discussed and generalized  in the context of Quantum Rotor model in the following sections.

\section{The Large \texorpdfstring{$n$}{n} limit of \texorpdfstring{$O(n)$}{O(n)} Quantum Rotors on Graphs}\label{section3}
The quantum rotor model is a paradigmatic model in the study of interacting quantum degrees of freedom. The rotors possess, in contrast to Heisenberg spins, a $O(n)$ symmetry and can be represented as $n$-component vectors living on a sphere \cite{sachdev2011quantum}:
\begin{equation}
    \hat{\mathbf{s}}_i^2=\sum_{\alpha=1}^n\hat{s}_{i,\alpha}^2=n,
\end{equation}
where the site index is $i$ and the components are denoted by $\alpha$. The normalization is again chosen in such a way that the large $n$ limit gives the quantum spherical model on regular lattices. The momenta $\hat{\mathbf{p}}_i$ associated to the rotors
fulfill canonical commutation relations:
\begin{equation}
    \left[\hat{s}_{i,\alpha},\hat{p}_{j,\beta}\right]=i\delta_{ij}\delta_{\alpha,\beta}.
\end{equation}
Placing the rotors on the vertices of a graph $\mathcal{G}$ the Hamiltonian of the quantum rotor model is defined as:
\begin{equation}
    H=-\sum_{i,j}A_{ij}\hat{\mathbf{s}}_i\cdot\hat{\mathbf{s}}_j+\frac{g}{2}\sum_i\hat{\mathbf{L}}^2_i,
\end{equation}
where the kinetic term has a coupling $g>0$ and is written in terms of the angular momentum operator $\hat{\mathbf{L}}$. 

The $O(n)$ rotor model has served as tool to study many aspects of quantum many body theory and it directly emerges in a variety of physical systems such as interacting bosons \cite{polak2007quantum}, Josephson junctions \cite{anglin2001exact} and magnetic systems \cite{chakravarty1989two}.

The large $n$ limit on a general graph can be taken by writing the partition function $Z$ as a functional integral over the action $S$:
\begin{align}
    Z&=\prod_i\int d{\mathbf{s}}_i(\tau) \delta(\mathbf{s}_i^2(\tau)-n)\exp(-S),\\
    S&=\int_0^\frac{1}{T}d\tau\left(\frac{1}{2g}\sum_i\left(\frac{\partial\mathbf{s}_i(\tau)}{\partial\tau}\right)^2-\sum_{ i,j}A_{ij}\mathbf{s}_i(\tau)\cdot\mathbf{s}_j(\tau)\right).
\end{align}
Using the integral representation of the delta function
\begin{equation}
    \delta(\mathbf{s}_i^2(\tau)-n)=\int^{i\infty}_{-i\infty} d\lambda_i(\tau)\,\exp\left[\lambda_i(\tau)(\mathbf{s}_i^2(\tau)-n)\right],
\end{equation}
one can write the action as the sum over $n$ independent components of the vectors $\mathbf{n}_i(\tau)$:
\begin{align}
    S&=\sum_{\alpha=1}^nS^{(\alpha)}\notag\\
    S^{(\alpha)}&=\int_0^\frac{1}{T}d\tau\left(\frac{1}{2g}\sum_i\left(\frac{\partial s_{i,\alpha}(\tau)}{\partial\tau}\right)^2\right.\notag\\   
    &-\left.\sum_{i,j}A_{ij}s_{i,\alpha}(\tau)s_{i,\alpha}(\tau)+\lambda_i(\tau)(s_{i,\alpha}^2(\tau)-1)\right).
\end{align}
In the large $n$ limit one can get replace the integration over $\lambda_i(\tau)$ by taking $\lambda_i(\tau)=\lambda_i$ as Lagrange multipliers enforcing the constraints $\langle n_i^2\rangle=1$ and one obtains an action in terms of scalar field $n_i$:
\begin{align}
    Z=&\int\prod_i ds_i(\tau)\exp(-S),\\
    S=&\int_0^\frac{1}{T}d\tau\left(\frac{1}{2g}\sum_i\left(\frac{\partial s_i(\tau)}{\partial\tau}\right)^2\right.\notag\\
    &\left.-\sum_{\langle ij\rangle}s_i(\tau)s_j(\tau)+\sum_i\lambda_i(s^{2}_i(\tau)-1)\right).\label{quantum_action}
\end{align}
Finally, one can write the Hamiltonian in operator form which reads:
\begin{equation}
    H=-\sum_{i,j}A_{ij}\hat{s}_i\hat{s}_j+\frac{g}{2}\sum_i\hat{p}^2_i+\sum_i\lambda_i(\hat{s}_i^2-1),\label{eq:quant_large_n}
\end{equation}
where $\hat{s}_i$ and $\hat{p}_i$ fulfill the commutation relations:
\begin{equation}
    [\hat{s}_i,\hat{s}_j]=[\hat{p}_i,\hat{p}_j]=0,\qquad [\hat{s}_i,\hat{p}_j]=i\delta_{ij}
\end{equation}
and $\lambda_i$ are determined by the constraint $\langle n_i^2\rangle=1$.

Notice that it differs from the quantum spherical model considered in \cite{vojta1996quantum} only by the fact that each site has its own Lagrange multiplier. On regular lattices it holds that $\lambda_i=\lambda$ for all $i$ and all constraints collapse to just one giving exactly the model defined in \cite{vojta1996quantum}.

Since the Hamiltonian describes coupled harmonic oscillators one can easily evaluate the free energy:
\begin{equation}
    f=-\frac{T}{N}\sum_k \ln \sinh\left( \sqrt{\frac{g \epsilon_k}{2}} \frac{1}{T}\right)+\frac{1}{2N}\sum_i(\lambda_i-g)
\end{equation}
where $N$ is the total number of sites and $\epsilon_k$ are the eigenvalues of the matrix M:
\begin{equation}
    M_{ij}=-A_{ij}+\delta_{ij}\lambda_i.
\end{equation}
The saddle point constraints are obtained by differentiation with respect to $\lambda_i$ and read:
\begin{equation}
    \frac{\partial f}{\partial\lambda_i}
    =-\frac{1}{2N}\sum_k |\psi_k(i)|^2
    \sqrt{\frac{g}{2\epsilon_k}}
    \coth\left(\sqrt{\frac{g \epsilon_k}{2}} \frac{1}{T}\right)+\frac{1}{2N}=0,\label{quantum_constraints}
\end{equation}
where $\psi_k(i)$ is the $i$-th component of the $k$-th eigenvector of $M$.
Taking the limit $g\rightarrow 0$ one approaches the result for the classical case \cite{titov2025laplacian} with an additional term $\sim \ln g/T$. The zero temperature limit gives an free energy of the form:
\begin{equation}
    f=-\frac{1}{N}\sqrt{\frac{g}{2}}\sum_k\sqrt{\epsilon_k}+\frac{1}{2N}\sum_i(\lambda_i-g).\label{eq:free_energy_zeroT}
\end{equation}
The corresponding saddle point constraints are:
\begin{equation}
    \frac{\partial f}{\partial\lambda_i}
    =-\frac{1}{2N}\sum_k |\psi_k(i)|^2
    \sqrt{\frac{g}{2\epsilon_k}}
+\frac{1}{2N}=0.\label{zeroT_constraints}
\end{equation}
In contrast to the classical large $n$ limit one can not write these as a determinant of $M$, but instead needs knowledge of all eigenvalues and eigenvectors. 

\section{Laplacian and Adjacency Matrix Limits}\label{section4}
For a general graph it is not feasible to determine the saddle point values of $\lambda_i$ by solving \eqref{quantum_constraints} analytically. It is, however, possible to evaluate them in different limits. 

For $g=0$, one obtains the result valid for the large $n$ limit of the classical $O(n)$ model on the same graph \cite{titov2025laplacian}:
\begin{equation}
    M_{ij}\sim \begin{cases}
			L_{ij}, & \text{for $T\rightarrow 0$},\\
            A_{ij}+\frac{1}{2T}\delta_{ij}, & \text{for $T\rightarrow \infty$}.
		 \end{cases}
\end{equation}
The low temperature behavior is dominated by the Laplacian matrix while the high temperature behavior is determined by the Adjacency matrix of the graph. In between these limits one has a crossover between the Laplacian 
(for small temperature) and the Adjacency matrix (for large temperature). If there is a phase transition then the singular part of the free energy at this point is determined by the Laplacian matrix.

\begin{figure*}
    \centering
    \includegraphics[width=0.8\linewidth]{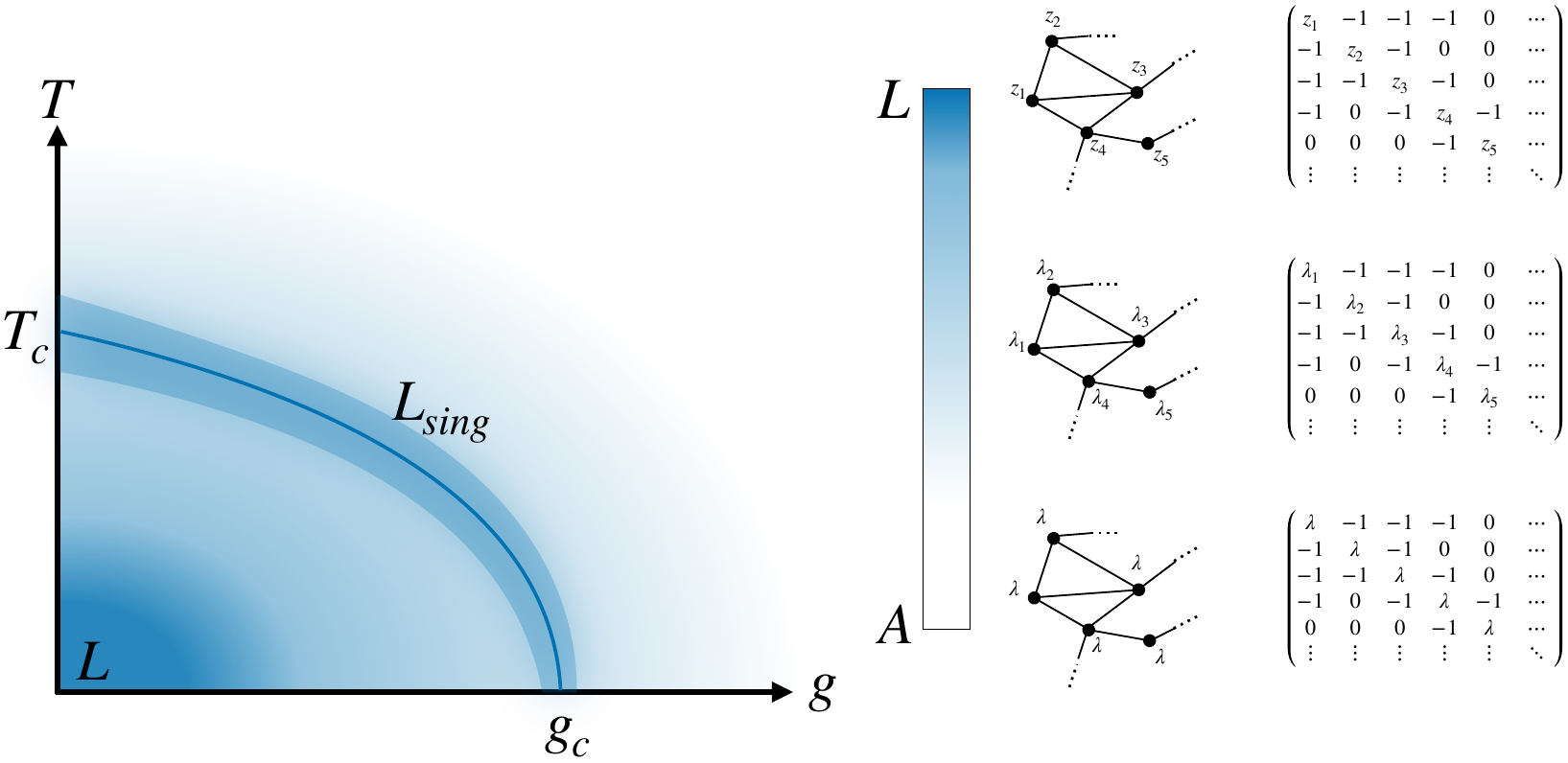}
    \caption{Qualitative phase diagram of the $O(n\to\infty)$ Quantum Rotor model on graphs with dimension $d_s>2$. The singular part of the free energy on the whole line is determined by the Laplacian matrix $L$, which we denote as $L_{\text{sing}}$. The quantum phase transition at $T=0$ has the dynamical critical exponent $z=1$. In the $T\to 0$ and $g\to 0$ limit the full free energy is governed by the Laplacian while for large values of $g$ and $T$ the Adjacency matrix $A$ controls the free energy.}
    \label{fig:1}
\end{figure*}

\begin{figure}
    \centering
    \includegraphics[width=\linewidth]{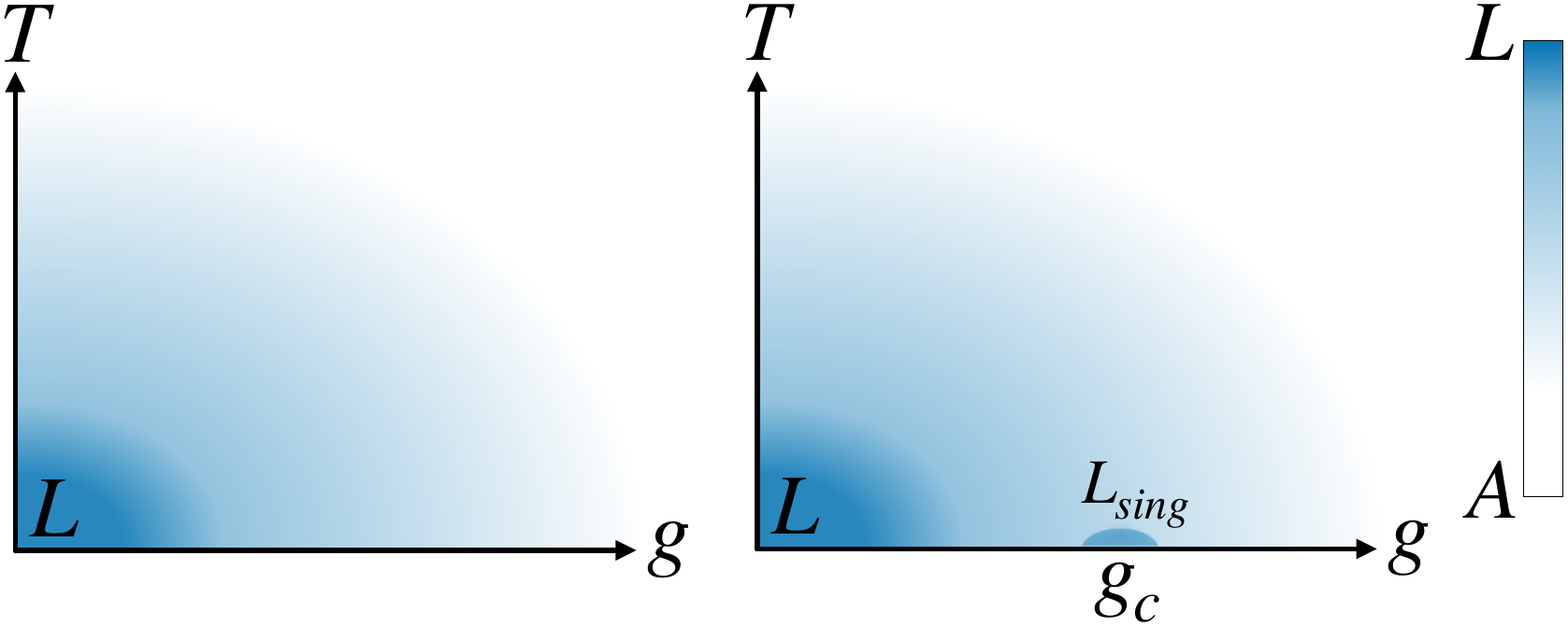}
    \caption{Qualitative phase diagram of the $O(n\to\infty)$ Quantum Rotor model on graphs with dimension $2\geq d_s>1$ (right) and $d_s\leq1$ (left). The darker region around the critical point $g_c$ only indicates that there the singular part is determined by the Laplacian matrix.}
    \label{fig:2}
\end{figure}

A similar analysis can be applied to $T=0$ and $g\to\infty$. By inspecting \eqref{zeroT_constraints}, one notices that in the large $g$ limit one must have $\epsilon_k\sim g$, from which $\lambda_i\approx C_i g$ follows, where $C_i$ are some constants. For a diagonally dominated matrix the eigenvectors are given by $\psi_k(i)\approx\delta_{ik}$ to leading order. In this limit the solutions of \eqref{zeroT_constraints} are therefore given by:
\begin{equation}
    M_{ij}\sim A_{ij}+\frac{g}{2}\delta_{ij},\quad \text{for $g\rightarrow \infty$}
\end{equation}
The last limiting case is to consider $T\to\infty$ and $g\to\infty$ while keeping $T/\sqrt{g}$ constant. One has to use \eqref{quantum_constraints}, but the same argument as before can be made and one finds:
\begin{equation}
    M_{ij}\sim A_{ij}+\frac{g}{2}\delta_{ij},\quad \text{for $g\rightarrow \infty$,\, $T\to\infty$.}
\end{equation}
Schematic phase diagrams are shown in Fig. \ref{fig:1} and Fig. \ref{fig:2} for the three cases: $d_s>2$, $2\geq d_s>1$ and $d_s\leq1$. For all of them,  the point $g=0$ and $T=0$ in the two dimensional parameter plane is where the full free energy is always determined by the Laplacian. Far away from this point, the Adjacency matrix is the relevant matrix. For $d_s\leq1$ there is no phase transition and one has just this crossover. In dimensions $2\geq d_s>1$ there is just quantum phase transition at $T=0$, whose critical exponents are determined by the Laplacian matrix as described in the following section. Finally, for $d_s>2$ there is a critical line, throughout which the Laplacian matrix determines the critical behavior.

Although these results paint a clear picture of the limiting behavior of the Lagrange multipliers, one cannot make concrete statements about how these limits are approached for a general graph, {\it i.e.}, the way in which these limits are reached is non-universal and depends on the specific graph. If one were to define a distance between the matrix $M$ and the Laplacian or Adjacency matrix then, in principle, it does not have to be monotonic in $T$ or $g$. In view of this, the goal of Fig. \ref{fig:1} and \ref{fig:2} is not to give a quantitative sketch of the distance, however defined, from the Laplacian or the Adjacency matrix, but to convey the idea that the system passes from $L$ to $A$ and that if there is a phase transition the Laplacian matrix is retrieved in the singular part. A study of how large this region around the critical line is, can be of interest for future works.

\section{Critical behavior}\label{section5}
In this section we analyze the critical behavior in two steps. First we show that the quantum rotor model in the large $n$ limit given by the Hamiltonian \eqref{eq:quant_large_n} and the accompanying constraints, has the same critical behavior as the quantum spherical model on graphs with the Hamiltonian:
\begin{equation}
    H=-\sum_{i,j}A_{ij}\hat{s}_i\hat{s}_j+\frac{g}{2}\sum_i\hat{p}^2_i+\lambda\sum_i(z_i\hat{s}_i^2-1),\label{eq:quant_spherical}
\end{equation}
with the constraint $\langle\sum_iz_iS_i^2\rangle=N$. Afterwards, we leverage this connection to determine the critical exponents associated with \eqref{eq:quant_spherical}, which is considerably easier to handle since it only has one constraint. 

\subsection{Relation to the Quantum Spherical Model}
First, we consider the finite temperature critical behavior. In this case one can expand the free energy \eqref{eq:quant_large_n} as a series:
\begin{align}
    F
    =&\frac{T}{2}\sum_k \ln \left( \epsilon_k\right)+T\sum_k \ln \left( \sqrt{\frac{g}{2}} \frac{1}{T}\right)+\\
    &T\sum_k \ln \left( 1+\left(\sqrt{\frac{g \epsilon_k}{2}} \frac{1}{T}\right)^2+\dots\right)+\frac{1}{2}\sum_i(\lambda_i-g).
\end{align}
Only the first term, which is the same as in the classical model can give a singular contribution and similarly for the free energy from \eqref{eq:quant_spherical}. Together with the equality of the singular parts of the classical free energy proven in \cite{burioni2000limit}, this immediately shows that the quantum models \eqref{eq:quant_spherical} and \eqref{eq:quant_large_n} have the same finite temperature critical exponents. As expected those exponents are equal to the classical ones.

For $T=0$ one cannot expand the free energy as a series and the techniques developed in \cite{burioni2000limit} cannot simply be applied to the resulting free energy \eqref{eq:free_energy_zeroT}.
To this end, we employ the classical to quantum mapping described in \cite{henkel1984hamiltonian} generalized to non translational invariant systems. Consider the two classical Hamiltonians defined on a graph $\mathcal{G}$ with Adjacency matrix $A\in \mathbb{R}^{N\times N}$ and spectral dimension $d_s$, augmented by an additional periodic direction at each site labeled by $\alpha$:
\begin{align}
    -\beta H_{\infty}&=\sum_{\alpha=1}^m\left[\mathbf{S}_{\alpha}^T(K_1A-\Lambda)\mathbf{S}_{\alpha}+K_2\mathbf{S}_{\alpha}^T\mathbf{S}_{\alpha+1}+\operatorname{Tr}\Lambda\right],\label{eq:quantclass1}\\
    -\beta H_{\text{spher}}&=\sum_{\alpha=1}^m\left[\mathbf{S}_{\alpha}^T(K_1A-\bar{\Lambda})\mathbf{S}_{\alpha}+K_2\mathbf{S}_{\alpha}^T\mathbf{S}_{\alpha+1}+\operatorname{Tr}\bar{\Lambda}\right]\label{eq:quantclass2},
\end{align}
The spins are now written as vectors $\mathbf{S}_{\alpha}=(S_{1,\alpha},\dots,S_{N,\alpha})$ and the matrices containing the Lagrange parameters are:
\begin{align}
    \Lambda&=\operatorname{diag}(\lambda_1,\dots,\lambda_N),\\
    \bar{\Lambda}&=\lambda\operatorname{diag}(z_1,\dots,z_N).
\end{align}
They do not depend on $\alpha$ because this direction is periodic: $\alpha_{N+1}=\alpha_1$. The anisotropic couplings $K_1$ and $K_2$ do not change the critical behavior compared to the isotropic case, as was shown in \cite{burioni1996universal}, which means \eqref{eq:quantclass1} and \eqref{eq:quantclass2} are in the same universality class. In the following, we perform the mapping for the large $n$ Hamiltonian \eqref{eq:quantclass1} only, as the mapping for the spherical one follows analogously.

First, write the partition function in terms of transfer matrices:
\begin{align}
    Z
    &=\int \prod_{\alpha=1}^m d\mathbf{S}_{\alpha}\,\exp\left(-\beta H_{\infty}\right)\\
    &=\int\prod_{\alpha=1}^m d\mathbf{S}_{\alpha}\, T(\mathbf{S}_{1},\mathbf{S}_{2})\dots T(\mathbf{S}_{m},\mathbf{S}_{1})=\operatorname{Tr}T^m,
\end{align}
where
\begin{align}
 T(\mathbf{S}_{\alpha},\mathbf{S}_{\alpha'})=&\exp\left(\frac{1}{2}\mathbf{S}_{\alpha}^T(K_1A-\Lambda)\mathbf{S}_{\alpha}\right.\\
 &\left.+\frac{1}{2}\mathbf{S}_{\alpha'}^T(K_1A-\Lambda)\mathbf{S}_{\alpha'}+K_2\mathbf{S}_{\alpha}^T\mathbf{S}_{\alpha'}\right).
\end{align}
To take the continuum limit in the "time" direction labeled by $\alpha$, one defines a lattice spacing $\tau$ in that direction and allows the coupling $K_1$ and $K_2$ to depend on the lattice spacing
Next, consider two wavefunctions $\psi$ and $\varphi$ and define:
\begin{align}
    \bar{\psi}(\mathbf{S}_{\alpha})=&\psi(\mathbf{S}_{\alpha})\exp\left(\frac{1}{2}\mathbf{s}_{\alpha}(K_1A-\Lambda)\mathbf{S}_{\alpha}+\frac{K_2}{2}\mathbf{S}_{\alpha}^2\right),\\
    \bar{\varphi}(\mathbf{S}_{\alpha})=&\varphi(\mathbf{S}_{\alpha})\exp\left(\frac{1}{2}\mathbf{s}_{\alpha}(K_1A-\Lambda)\mathbf{S}_{\alpha}+\frac{K_2}{2}\mathbf{S}_{\alpha}^2\right),
\end{align}
which is done to simplify the calculation of the expectation value $\langle \psi \vert T\vert \varphi\rangle$:
\begin{equation}
    \langle \psi|T|\varphi\rangle =\int d\mathbf{S}_{\alpha}d\mathbf{S}_{\alpha'}\bar{\psi}(\mathbf{S}_{\alpha})\exp\left(-\frac{K_2}{2}(\mathbf{S}_{\alpha}-\mathbf{S}_{\alpha'})^2\right)\bar{\varphi}(\mathbf{S}_{\alpha'}).
\end{equation}
For small $\tau$ one obtains the standard Gaussian kernel expansion for the transfer matrix:
\begin{align}
    T(\mathbf{S}_{\alpha},\mathbf{S}_{\alpha'})=&\left(\frac{2\pi}{K_2}\right)^{m/2}\delta(\mathbf{S}_{\alpha}-\mathbf{S}_{\alpha'})\left(1+\mathbf{S}_{\alpha}^T(K_1A-\Lambda)\mathbf{S}_{\alpha}\right.\notag\\
    &\left.-K_2\mathbf{S}_{\alpha}^2+\frac{1}{2K_2}\Delta+\dots\right),
\end{align}
where $\Delta=\sum_{i}\partial^2/\partial S_{i,\alpha}^2$. Rescaling the coupling parameters and the Lagrange multipliers in the following way:
\begin{align}
    K_1&=K'_1\tau,\\
    K_2&=K'_2/\tau,\\
    \lambda_i-K_2&=\lambda'_i\tau,
\end{align}
allows one to write the transfer matrix in the $\tau\to0$ limit as:
\begin{align}
    T(\mathbf{S}_{\alpha},\mathbf{S}_{\alpha'})=&\left(\frac{2\pi}{K_2}\right)^{m/2}\delta(\mathbf{S}_{\alpha}-\mathbf{S}_{\alpha'})\\
    &\times\left[1+\tau\left(\mathbf{S}_{\alpha}^T(K'_1A-\Lambda')\mathbf{S}_{\alpha}+\frac{1}{2K'_2}\Delta+\dots\right)\right].
\end{align}
The Hamiltonian is therefore given by:
\begin{equation}
    H=\mathbf{S}(K'_1A-\Lambda')\mathbf{S}+\frac{1}{2K'_2}\Delta,
\end{equation}
which reads in operator form:
\begin{equation}
    \hat{H}=K_1'\sum_{i,j}\hat{s}_iA_{ij}\hat{s}_j+\frac{1}{2K'_2}\sum_{i}\hat{p}^2_i-\sum_i\lambda'_i\hat{s}_i^2.
\end{equation}
This is just the Hamiltonian of the $O(n)$ rotor model in the large $n$ limit and starting from \eqref{eq:quantclass2} one would end up with the Hamiltonian of quantum spherical model. One therefore has shown that starting from two classical models in the same universality class, one is mapped to the $O(n)$ rotor model in the large $n$ limit and the other to the quantum spherical model. This concludes the proof that both models share the same critical behavior. 

\subsection{Critical exponents}
Having shown that \eqref{eq:quant_large_n} and \eqref{eq:quant_spherical} have the same critical exponents, one is able to consider just the much simpler Hamiltonian \eqref{eq:quant_spherical} in this section. 
We write the action of the spherical model on a graph with Adjacency matrix $A$:
\begin{align}
    S=&\int_0^\frac{1}{T}d\tau\left(\frac{1}{2g}\sum_i\left(\frac{\partial n_i(\tau)}{\partial\tau}\right)^2\right.\notag\\
    &\left.-\sum_{i,j}n_i(\tau)(A_{ij}-\lambda z_i\delta_{ij})n_j(\tau)\right),
\end{align}
with the constraint $\langle\sum_iz_iS_i^2\rangle=N$.
Next, one introduces the Matsubara frequencies $\omega_n=2\pi m T$:
\begin{equation}
    S=\sum_{m=-\infty}^\infty \left(-\sum_{i,j}n_i(\omega_m)(A_{ij}-\frac{\omega_n^2}{2g}-\lambda z_i\delta_{ij})n_j(\omega_m)\right),
\end{equation}
which allows one to obtain the free energy:
\begin{equation}
    f=-\frac{1}{N}\sum_{m=-\infty}^\infty\operatorname{Tr}\ln\left(-A+\frac{\omega_n^2}{2g}\mathbb{I}+\lambda Z\right)+N\lambda+const.
\end{equation}
where $Z=\operatorname{diag}(z_1,\dots,z_{\mathcal{N}})$. The spherical constraint $\partial f/\partial\lambda=0$ gives rise to the saddle point equation:
\begin{equation}
    \frac{1}{N}\sum_{m=-\infty}^\infty\sum_i z_i\left(-A+\frac{\omega_n^2}{2g}\mathbb{I}+\lambda Z\right)^{-1}_{ii}=1.\label{eq:matsubara_constraint}
\end{equation}
At zero temperature the sum turns into an integral and one ends up with the constraint:
\begin{equation}
    \frac{1}{N}\int_{-\infty}^\infty d\omega\sum_i z_i\left(-A+\frac{\omega^2}{2g}\mathbb{I}+\lambda Z\right)^{-1}_{ii}=1.
\end{equation}
The existence of a phase transition depends on the existence of real solutions to the above equation. It has been shown that the matrix has the following singular behavior \cite{burioni1996universal}:
\begin{align}
    \lim_{s\rightarrow 0^+}\operatorname{sing}\frac{1}{N}\sum_i\left(-A+Z+sM\right)^{-1}_{ii}\sim s^{d_s/2-1},
\end{align}
where $M=\operatorname{diag}(m_1,\dots,m_{\mathcal{N}})$ is any distribution of finite masses $m_i$. Expanding around $\omega\approx0$ and $\lambda\approx\lambda_c=1$ one can extract the singular part:
\begin{align}
    &\operatorname{sing}\frac{1}{\mathcal{N}}\int_{-\infty}^\infty\sum_i z_i\left(-A+\frac{\omega^2}{2g}\mathbb{I}+\lambda Z\right)^{-1}_{ii},\notag\\
    &\approx\frac{1}{\mathcal{N}}\int_{-\infty}^\infty\sum_i (\lambda-1+\omega^2/2g)^{d_s/2-1}
    \sim (\lambda-1)^{(d_s-1)/2}.
\end{align}
All critical exponents can be directly obtained now in the usual way for the spherical model and are equal to the ones of the $d_s+1$ dimensional classical model. 

\section{Conclusions}
In this paper we have shown that the quantum $O(n)$ rotor model in the large $n$ limit on general graphs is always governed by the Laplacian in the small temperature and quantum coupling limit, while the Adjacency matrix determines the behavior far away from this point in the parameter plane. This interplay emerges through the Lagrange multipiers, which enforce the spin constraints. Although their number diverges in the thermodynamic limit they simplify for either large or small values of temperature and coupling constant. This gives a direct connection between the two central graph-theoretic quantities:.

We have also analyzed the critical behavior and have shown that it is fully determined by the spectral dimension and is equivalent to the quantum spherical model on the same graph. To this end, we used known results for the classical case and generalized them to the quantum model, by using the functional integral representation and a classical to quantum mapping. 

Our work can serve as a starting point for a systematic $1/n$ expansion on graphs and networks relevant for applications in ultracold atoms and other spin models. The, in some sense, nonphysical global constraint of the spherical model on graphs is replaced by local constraints that are the product of a large $n$ limit. Since the model remains Gaussian, one can study dynamical properties in a direct way, both numerically and analytically. A time dependent coupling $g(t)$ or a non equilibrium initial state lead to a time evolution of $\lambda_i(t)$, which we think is an interesting subject of future work.

\begin{acknowledgements}
    Discussions with N. Defenu are gratefully acknowledged. The authors acknowledge funding from the European Union’s Horizon Europe Research and Innovation Programme under the Marie Sk\l{}odowska-Curie Doctoral Network MAWI (Matter-Wave Interferometers) under the grant agreement No. 101073088.
\end{acknowledgements} 

\bibliography{bibliography}
\end{document}